# Control of Two-Dimensional Excitonic Light Emission *via* Photonic Crystal

Sanfeng Wu[1], Sonia Buckley[2], Aaron M. Jones[1], Jason S. Ross[3], Nirmal J. Ghimire[4,5], Jiaqiang Yan[5,6], David G. Mandrus[4,5,6], Wang Yao[7], Fariba Hatami[8], Jelena Vučković[2], Arka Majumdar[9], Xiaodong Xu[1,3,*]

[1]Department of Physics, University of Washington, Seattle, Washington 98195, USA
[2]Ginzton Laboratory, Stanford University, Stanford, CA 94305, USA
[3]Department of Material Science and Engineering, University of Washington, Seattle, Washington 98195, USA
[4]Department of Physics and Astronomy, University of Tennessee, Knoxville, Tennessee 37996, USA
[5]Materials Science and Technology Division, Oak Ridge National Laboratory, Oak Ridge, Tennessee, 37831, USA
[6]Department of Materials Science and Engineering, University of Tennessee, Knoxville, Tennessee, 37996, USA
[7]Department of Physics and Center of Theoretical and Computational Physics, University of Hong Kong, Hong Kong, China
[8]Department of Physics, Humboldt University, D-12489, Berlin, Germany
[9]Department of Electrical Engineering, University of Washington, Seattle, Washington 98195, USA

*Correspondence to: xuxd@uw.edu

**Monolayers of transition metal dichalcogenides (TMDCs) have emerged as new optoelectronic materials in the two dimensional (2D) limit, exhibiting rich spin-valley interplays, tunable excitonic effects, and strong light-matter interactions. An essential yet undeveloped ingredient for many photonic applications is the manipulation of its light emission. Here we demonstrate the control of excitonic light emission from monolayer tungsten diselenide ($WSe_2$) in an integrated photonic structure, achieved by transferring one monolayer onto a photonic crystal (PhC) with a cavity. In addition to the observation of greatly enhanced (~60 times) photoluminescence of $WSe_2$ and an effectively coupled cavity-mode emission, we are able to redistribute the emitted photons both polarly and azimuthally in the far field through designing PhC structures, as revealed by momentum-resolved microscopy. A 2D optical antenna is thus constructed. Our work suggests a new way of manipulating photons in hybrid 2D photonics, important for future energy efficient optoelectronics and 2D nano-lasers.**

**Text**

Major efforts in optoelectronic research focus on searching for the proper materials and designs for critical components, such as light emitters, optical modulators, converters and detectors. Monolayers of TMDCs, i.e., $MX_2$ (M=Mo, W; X=S, Se, Te), have been regarded as promising candidates for these applications[1], thanks to their outstanding semiconducting behaviors in the 2D limit[2-10] and exceptional ability to convert light into photo-current[11] in an atomically thin structure. Based on these properties, single layer $MoS_2$ transistors[12], ultrahigh-gain phototransistors[13], ultrasensitive photodetectors[14] and light-emitting diodes (LED) [15-18] have already been demonstrated.

Many of the novelties and potential applications of these monolayer TMDCs lie in their excitonic light emissions. A controllable and directional emission in such systems is thus highly desired for developing efficient photonic and optoelectronic components. PhCs, periodic optical nanostructures, and photonic crystal cavities (PhCCs) are powerful platforms for manipulating light emission[20], guiding on-chip photons[19], and enhancing light-matter interactions[21-22]. A successful integration of 2D TMDCs with PhC structures would provide a powerful way to manipulate their exotic excitonic emissions, as well as represent a novel optoelectronic hybrid capable of harnessing the advantages of both monolayer semiconductors and PhCs.

The idea of utilizing PhCs to modify light emission and enhance quantum yield from a semiconducting emitter has led to fruitful achievements in quantum well LEDs[20, 23, 24]. Figure 1a and 1b show typical devices with light emitting materials (multi-quantum wells) embedded into a 2D PhC, which employs its photonic band-gap[23] and diffraction grating effect[24], to redistribute the emitted photons. However, in terms of being compatible with integrated electronic circuits[20], these designs have potential difficulties: (1) it is challenging to fabricate contacts[25] transverse to the light-active material to form electronic elements like transistors; (2) light extraction is limited to the light cone from which light can escape total internal reflection at the air-semiconductor surface; (3) the low-order guide modes in the diffraction grating approach (Fig. 1b) interact poorly with the PhCs, limiting the device performance; (4) a considerable amount (~30%) of the embedded quantum wells are etched away during PhC fabrication (Fig. 1a), also leading to strong non-radiative surface recombination and degradation of the excitonic properties of the quantum well as a result of the fabrication process[20].

In this Letter, we present integrated monolayer TMDC/PhC and PhCC photonic devices that overcome the difficulties mentioned above with great control over photon emission. Figure 1c shows our design based on a monolayer TMDC, which is placed on top of a 2D photonic crystal. This geometry provides the following advantages. First, the entire monolayer sheet is open for fabrication of both contacts and gates, allowing for further development of optoelectronic devices. Second, unlike designs which embed the light-active material, there is no loss channel due to total internal reflection and therefore the light cone is the whole upper hemisphere[20]. Third, since the thickness of the semiconducting layer is pushed to the atomic limit, the guided mode is now restricted to the vicinity of the PhC where interactions are strong[20]. Fourth, instead of being destroyed during etching, the monolayer TMDC is placed after PhC fabrication, which preserves the high quality of the emitting layer. These benefits are unprecedented in conventional structures with embedded light emitters.

Monolayer $WSe_2$ was selected as our light emitting material since it exhibits strong excitonic photoluminescence (PL) (e.g. more than 40 times higher than single layer $MoS_2$)[26]. To fabricate the hybrid structure, a single layer of $WSe_2$ was mechanically exfoliated onto a PMMA layer and then transferred to the PhC by standard methods recently developed in 2D material research[27]. Figures 1d and e show the optical and scanning electron micrograph (SEM) of a typical device (device A). The photonic crystal is fabricated from a 180nm thick gallium phosphide (GaP) membrane on top of a 800nm AlGaP sacrificial layer sitting on bulk GaP[28]. A triangular lattice of holes is fabricated in the GaP membrane (see Methods) and a PhCC is defined by omitting holes in the lattice. We choose the lattice parameters so that the PhCC modes match the exciton emission from monolayer $WSe_2$. The hole radius, $r$, is about 50nm and the lattice constant, $a$, is 200nm for devices A and C with a linear four-hole defect, L4. For device B, $r = 100$nm, $a = 375$nm with L3. For devices A and C the exciton emission is within the bandgap of the photonic crystal cavity, while for device B the exciton emission is outside the photonic crystal bandgap. After removing the sacrificial layer, the hole depth, $D$, from the top surface of the PhC to the bottom of the GaP bulk, is estimated to be ~980nm for all devices.

We then study the device via PL measurements pumped by a 660nm laser focused with a 100x objective (NA=0.95, see methods). Figure 2a shows a 2D scanning PL intensity map of device A, where the dashed white line indicates the PhC area. The corresponding sample area is illustrated in Fig. 2b. The rectangular shape of the PhC is well-mapped in the PL which clearly demonstrates that the monolayer on the PhC yields remarkably stronger PL than off the crystal. We selectively plot three spectra taken from off-crystal, on-crystal and on-cavity (circles in Fig. 2b) together in Fig. 2c. The total PL intensity on PhC is enhanced 20 times compared to that of off PhC. The highest enhancement ratio obtained among all our devices is about 60 times (see supplementary materials). PL intensity of TMDC monolayers is known to strongly depend on the substrate[2] and the surrounding molecular environment[29]. Our observation indicates that PhCs significantly increase the radiative recombination of excitons in $WSe_2$. The photonic band-gap effect, which inhibits the spontaneous emission rate laterally and redistributes the light emission into the vertical direction, may play a role in this device since the light emission falls into the forbidden region[23].

The effective coupling between monolayer and PhC is further revealed by the resonant peak emission at 756nm (red spectrum in Fig. 2c) when a cavity is present. The presence of this peak indicates coupling to the cavity mode. To clearly show this, we map out the peak amplitude difference between emissions at 756nm (on cavity mode) and 748nm (off cavity mode) in Fig. 2d. We can see that such peak emission is strongly restricted to the cavity location with ~1μm lateral size, approximately the length of the cavity. Moreover, PL emission from the cavity is highly polarized in the $x$ direction, as shown in Fig. 2e. We found that this polarized PL is independent of the polarization of the excitation beam. Such polarization is defined by the corresponding cavity mode. We identify the cavity mode as the $x$-polarized dipole-like mode of the L4 cavity. This mode has a Q factor of ~250 which is sufficient for matching to the broad exciton PL, and strongly radiates in the vertical direction. The Q factor changes to ~180 in excitonic emission after $WSe_2$ deposition. All the observed modes are identified by a quantitative comparison with the FDTD simulations of the cavity[30] (supplementary materials).

By varying the lattice parameters of the PhCC, we are able to tune the emission energy of the polarized PL. To investigate this, device C is designed to have similar Q factor but different lattice parameters (slightly larger hole radius). The PhCCs were characterized by cross-polarized reflectivity measurements[28] (horizontal incident, vertical detection, sample at 45°) before $WSe_2$ transfer, as shown in the top of Fig. 2f. This clearly shows a shift in resonant mode energy with lattice parameters. We get good agreement between this simulation and the cross-polarized reflectivity measurement for the variety of lattice constants fabricated for all the modes. The bottom of Fig. 2f plots the polarized PL (x-polarized emission minus y-polarized emission) from the WSe$_2$ on two different PhCCs (devices A and C). The PL emission unambiguously follows the cavity modes. The mechanism for this peak emission can be attributed to the Purcell effect, by which the spontaneous emission rate of the monolayer is increased[31]. In our case, the peak enhancement of the emission rate, *i.e.*, maximum Purcell factor, can be estimated to be $\frac{3}{4\pi^2}\frac{Q}{V}(\frac{\lambda_c}{n})^3 \sim 14$, where $\lambda_c \sim 756 nm$ is the wavelength of cavity emission, $n \sim 3.1$ is the refractive index of GaP, and $V \sim (\frac{\lambda_c}{n})^3$ is the mode volume. This enhancement, together with the inhibition from the PhC effect, leads to the resonance in the collected PL emission[32].

Based on the effective coupling between monolayer WSe$_2$ and PhC, we further investigate the far field distribution of light emission from our devices. We find that the emission pattern can be significantly modified by the PhC, through the diffraction grating effect. Momentum resolved microscopy (see methods) is applied to obtain the in-plane ($k_x$, $k_y$) momentum distribution of the far field light emission[33]. The in-plane momentum is related to the emission polar angle $\theta$ (Fig. 3) by $\sin(\theta) = k_\parallel/k_0$, where $k_\parallel = |\mathbf{k}_x + \mathbf{k}_y|$, and $k_0 = 2\pi/\lambda$ is the amplitude of the wave vector of the emitted photon with wavelength $\lambda$.

Figure 3a plots the normalized PL momentum distribution for samples placed on top of an unpatterned area (off crystal) on the GaP substrate of device A. We also present the data from WSe$_2$ on SiO$_2$ substrate in the supplementary materials. There is a subtle difference between the two emission patterns. On SiO$_2$ the light is more vertically directed. Despite these subtleties, both patterns show overall similar behavior with intensity decaying from the center. However, the pattern remarkably changes when the monolayer is on a PhC. Shown in Fig. 3b, the exciton emission is highly directed in a ring with polar angle $34^o < \theta < 48^o$ (defined in Fig. 3c), corresponding to a strong localization of in-plane momentums between $0.55k_0$ and $0.75k_0$. Figure 3d compares emission intensities from the different substrates versus polar angle $\theta$ along the fixed line $k_x = 0$. For comparison, a Lambertian emission is also presented as the green line. We can see that the light emission from planar substrates tends to have Lambertian-like behavior. However, the PhC drives the emission pattern greatly away from Lambertian to act like an optical antenna[34].

We attribute such antenna-like behavior to the diffraction grating effect of the PhC. Light emission from the periodic pattern interferes in the far field, where the photons coming from different holes experience differing optical path lengths (Fig. 4a). We simulate this effect by considering a 2D triangular diffraction grating. Since the lattice constant $a$ is smaller than $\lambda$, the diffraction equation $\sin\theta = m\lambda/a$ only holds for $m = 0$. Considering the etch depth of $D \sim 5\lambda/4$, photons reflected from the bottom surface destructively interfere with those from the top surface, thus suppressing vertical light emission, which corresponds to the primary (1$^{st}$) maximum of the zeroth order diffraction pattern. Therefore, the secondary (2$^{nd}$) maximum dominates the diffraction. The 3$^{rd}$ maximum also shows up in experiment, as indicated in Fig. 3d. The simulated result is presented in Fig. 3e, where experimental parameters are used ($a = 200nm$, $\lambda = 750nm$, NA=0.95). The corresponding grating pattern used here is shown in the supplementary materials. Our simulation agrees well with the experimental observation, reproducing the antenna pattern and also the emission angle for both 2$^{nd}$ and 3$^{rd}$ diffraction maxima.

In order to demonstrate that the directional emission can be controlled by designing the PhC, we perform a similar measurement on device B where the lattice constant is increased to $375nm$. We found that the emission can not only be directed to certain polar angles, but also in the azimuthal angle, as shown in Fig. 4b. Instead of being uniformly distributed along the azimuthal angle, the emission in this device shows interesting islands for both the 2$^{nd}$ (4 islands labeled by Γ) and 3$^{rd}$ (6 islands labeled by Λ) maxima. The change of the grating pattern (supplementary material) caused by increasing lattice constant is responsible for the PL pattern. The theoretical simulation, presented in Fig. 4c without any free parameters, is able to capture most of the detailed features observed in the experiment.

In summary, our experimental observations show great promise for exploring exotic phenomena and future novel applications based on 2D TMDC/PhC hybrids. Further fabrication of two top gates can lead to an atomically thin LED integrated with a PhCC, which will lead to interesting nano-photonic devices such as a single mode LED, potentially with spin or valley polarization, and possible 2D nano-lasers using a high quality factor cavity. The high directionality of the emission also suggests that such hybrid structures may be useful for highly efficient photon sources. The possibility of integration of electronics and photonics in these devices may additionally open up applications in on-chip communications, sensing, and optical computing.

**Methods**

**Device Fabrication**:

A 180 nm GaP membrane was fabricated on top of a 0.8 μm sacrificial layer on a (001)-oriented GaP wafer using gas-source molecular beam epitaxy (GSMBE). A layer of ~330 nm electron beam lithography resist (ZEP 520a) was spun on top of the wafer. The pattern was defined using a base dose range of 250-375 μC/cm$^2$ using a 100 keV electron beam lithography tool (JEOL JBX 6300). Following development, the pattern was etched deep into the membrane and sacrificial layer using a $Cl_2$ and $BCl_3$ chemistry in a plasma etching system. The resist was removed using microchem remover-PG. The sacrificial layer was then removed using a 7% HF wet etch.

Monolayer $WSe_2$ was transferred on top of PhC by standard procedures. We first spin-coated PVA (1%) and PMMA (950, 6%) on silicon chip followed by baking the chip at 180 $^o$C for 1 minute. After we exfoliated the $WSe_2$ flakes onto the PMMA/PVA/Si stacking substrate, The $WSe_2$/PMMA membrane was then separated from Si chip by dissolving PVA layer in water. The membrane was scoped up by a loop and the monolayer flake was precisely placed onto the PhC region under microscope, forming a PMMA/monolayer-$WSe_2$/PhC structure after heating. PMMA layer was removed by a 2-hour acetone bath.

**Momentum resolved microscopy**:

The momentum resolved microscopy was performed by imaging the back focal plane of the objective lens (Olympus MS Plan 100X/0.95/IC100) which was used for both focusing the laser beam onto sample and collecting the PL signal. The back focal plane image was focused to the open slit of the ANDOR spectrometer by a Bertrand lens with focal length of 150mm after a telescope setup to optimize the size of the image. The image was then collected by ANDOR newton CCD (1024 x 255 active pixels) by setting the spectrometer grating to zero. The normal PL measurements can be easily coupled to this system. The principle of this momentum resolved measurements can be found in Ref. 33.

**Acknowledgments:** The authors would like to thank Sen Yang for suggestions in momentum resolved measurements and Grant Aivazian for proof-reading. This work was mainly supported by DoE, BES, Materials Science and Engineering Division (DE-SC0008145). PhC fabrication was performed in part at the Stanford Nanofabrication Facility of NNIN supported by the National Science Foundation under Grant No. ECS-9731293, and at the Stanford Nano Center. SB was supported by a Stanford Graduate Fellowship. SB and JV were also supported by the Presidential Early Award for Scientists and Engineers (PECASE) administered through the Office of Naval Research, under grant number N00014-08-1-0561. NJG, JY, and DGM were supported by US DoE, BES, Materials Sciences and Engineering Division.

**Author Contributions**:

XX and AM conceived the experiments. SB and AM fabricated and characterized PhCs under JV's supervision. SW prepared and transferred monolayer $WSe_2$ samples, and performed the measurements with assistance from AMJ. SW analyzed the data and did simulation with discussion from SB, AM, WY, JV and XX. NJG, JY and DGM provided the bulk $WSe_2$. FH grew the GaP membrane. JSR performed SEM. SW wrote the paper with input from all authors.

**Figure Legends:**

**Figure 1 | Hybrid Monolayer $WSe_2$/PhC nano-structure. a** and **b,** Two types of conventional structures coupling PhC and a light emitting material (green layer), employing photonic band gap effect and diffraction grating effect, respectively. Both structures embed the light active material. The arrows in (b) denote the loss channels from total internal reflection and the low order guided mode. **c,** Schematic of coupled TMDC/PhC structure. **d,** Optical microscope image of device A. The green area is the PhC and the yellow area is $WSe_2$ bulk. The monolayer is within the orange dashed line. Scale bar: $5\mu m$. **e,** SEM image of the area of interest (indicated by blue dashed line in **d**). Scale bar: $3\mu m$.

**Figure 2 | Photoluminescence characterization of the Hybrid Photonic Structures. a,** 2D photoluminescence (PL) intensity color map of device A. **b,** The corresponding SEM image of the PhC area with defined axes. **c,** Spatially selected PL of device A. The red, black and green spectra are respectively taken from on-PhCC, on-PhC and off-PhC excitation, indicated by the circles in **b. d,** 2D map of peak height difference between PL at 756nm (on cavity resonance)

and 748nm (off cavity resonance), showing the cavity resonance and mapping out the cavity region. **e,** Linear-polarization resolved PL spectra taken for on-cavity excitation, depicting that the corresponding cavity-mode emission is polarized in *x* direction. **f,** Cross-polarized reflection spectra of two devices *v.s.* polarized PL spectra, demonstrating the control over the polarized light emission by varying the cavity parameter. Device A and C both have a lattice constant of 200 nm, with device C having a slightly larger hole radius, leading to a blueshift of the cavity resonance.

**Figure 3 | Control of excitonic light emission from monolayer WSe$_2$ *via* PhC. a,** Normalized photoluminescence intensity distribution over in-plane momentum space measured by momentum resolved microscopy, when the monolayer is placed on GaP substrate (off crystal area in device A). **b,** The same map for sample placed on top of PhC, showing a strikingly different distribution of the light emission. **c,** Cartoon plot defining the polar angle of the emission. **d,** Polar plot of the emission intensities of the different substrates with $k_x = 0$. As a comparison, the Lambertian type of emission is also plotted. The emission from PhC exhibits an optical-antenna behavior while planar GaP substrate behaves much like Lambertian. **e,** Theoretical plot of the momentum distribution of the emission predicted by diffraction grating effect, which agrees very well with the observation.

**Figure 4 | Control of both polar and azimuthal angle of WSe$_2$ photon emission. a,** Diffraction grating effect and the emission intensity image on the back focal plane of the objective lens. **b,** Normalized photoluminescence momentum map for device B with lattice constant $a = 375nm$, showing that not only the emission over polar angle, but also the azimuthal angle can be redistributed by PhC. **c,** Simulation of the emission pattern of device B.

# Figure 1

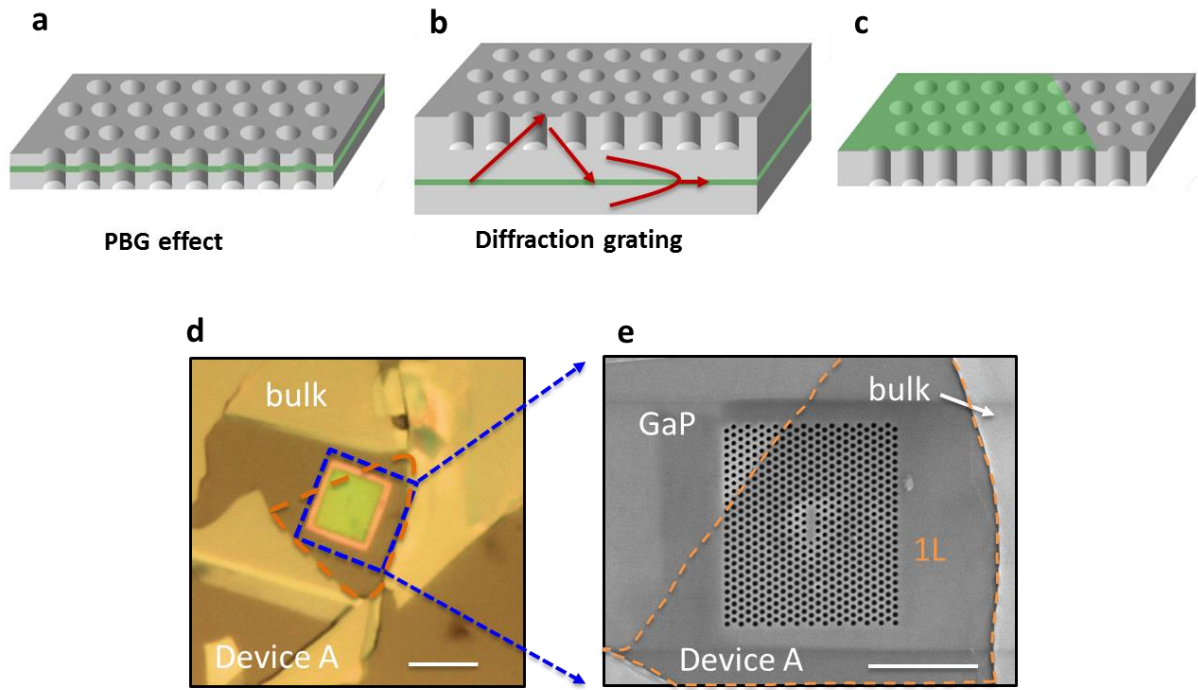

# Figure 2

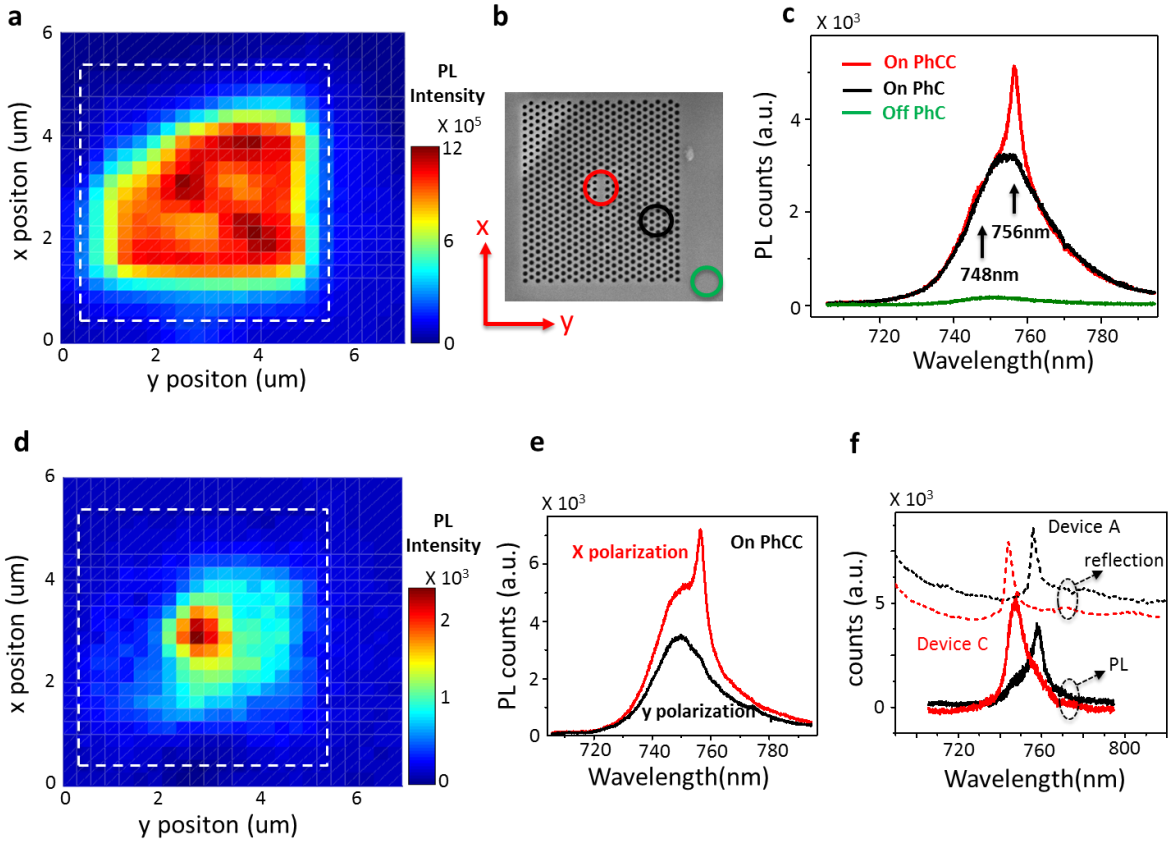

# Figure 3

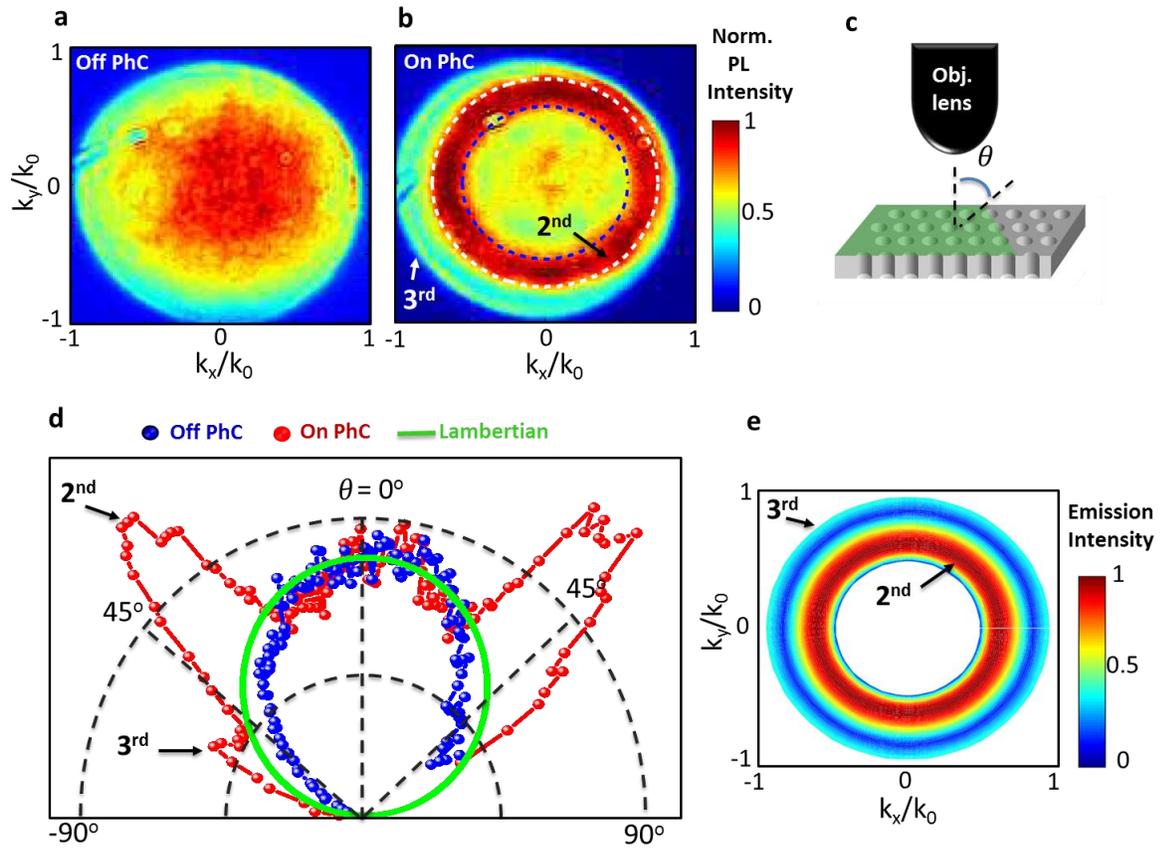

# Figure 4

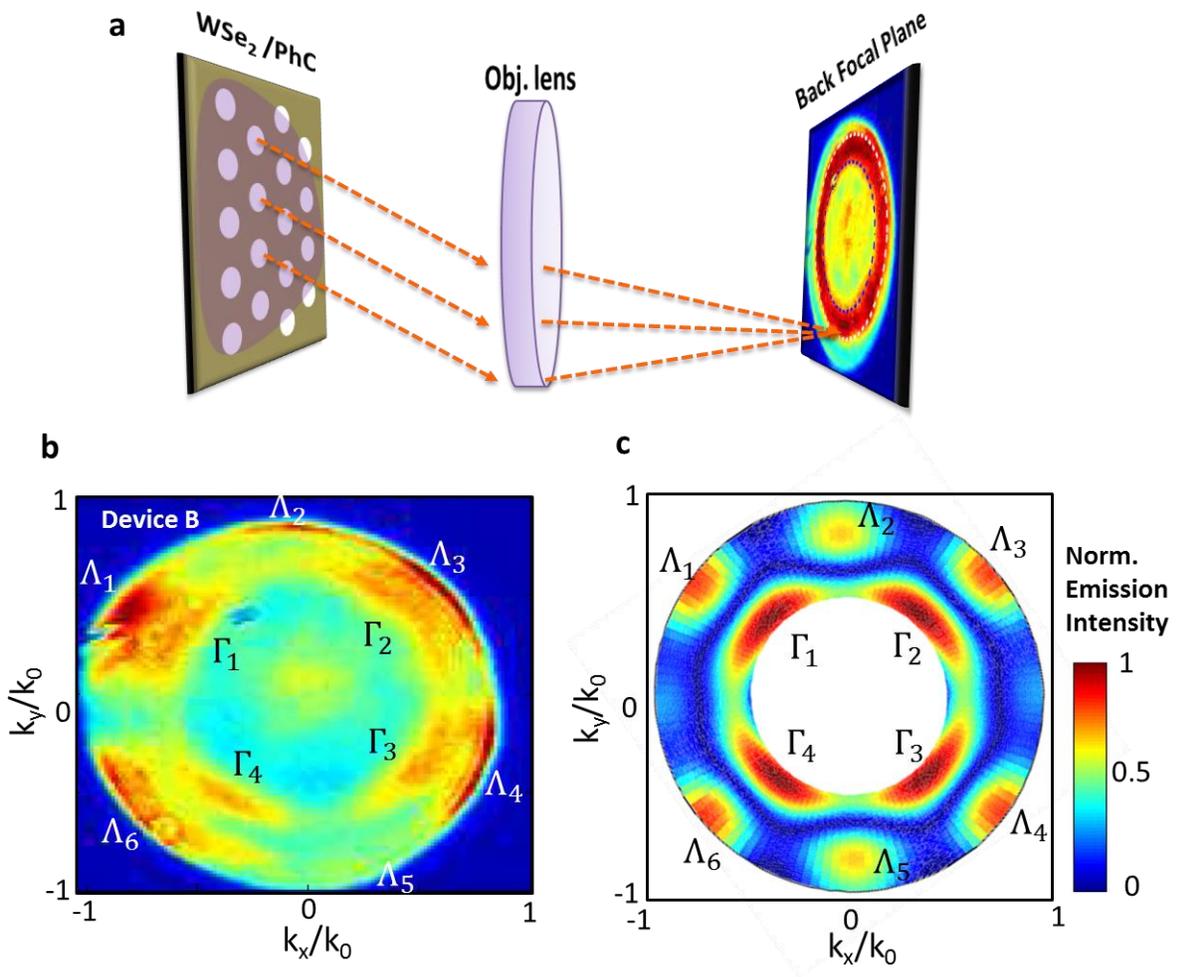

# Control of Two-Dimensional Excitonic Light Emission *via* Photonic Crystal

Sanfeng Wu[1], Sonia Buckley[2], Aaron M. Jones[1], Jason S. Ross[3], Nirmal J. Ghimire[4,5], Jiaqiang Yan[5,6], David G. Mandrus[4,5,6], Wang Yao[7], Fariba Hatami[8], Jelena Vučković[2], Arka Majumdar[9], Xiaodong Xu[1,3]

[1]Department of Physics, University of Washington, Seattle, Washington 98195, USA
[2]Ginzton Laboratory Stanford University, Stanford, CA 94305-4088 USA
[3]Department of Material Science and Engineering, University of Washington, Seattle, Washington 98195, USA
[4]Department of Physics and Astronomy, University of Tennessee, Knoxville, Tennessee 37996, USA
[5]Materials Science and Technology Division, Oak Ridge National Laboratory, Oak Ridge, Tennessee, 37831, USA
[6]Department of Materials Science and Engineering, University of Tennessee, Knoxville, Tennessee, 37996, USA
[7]Department of Physics and Center of Theoretical and Computational Physics, University of Hong Kong, Hong Kong, China
[8]Department of Physics, Humboldt University, D-12489, Berlin, Germany
[9]Department of Electrical Engineering, University of Washington, Seattle, Washington 98195, USA

**S1. Spectra of the device exhibiting 60x enhancement of photoluminescence**

**S2. Momentum resolved photoluminescence of monolayer $WSe_2$ on $SiO_2$ substrate**

**S3. Simulation of the emission distribution based on the diffraction grating effect**

**S4. Cavity mode profile simulation**

**S1. Spectra of the device exhibiting 60x enhancement of PL quantum yield**

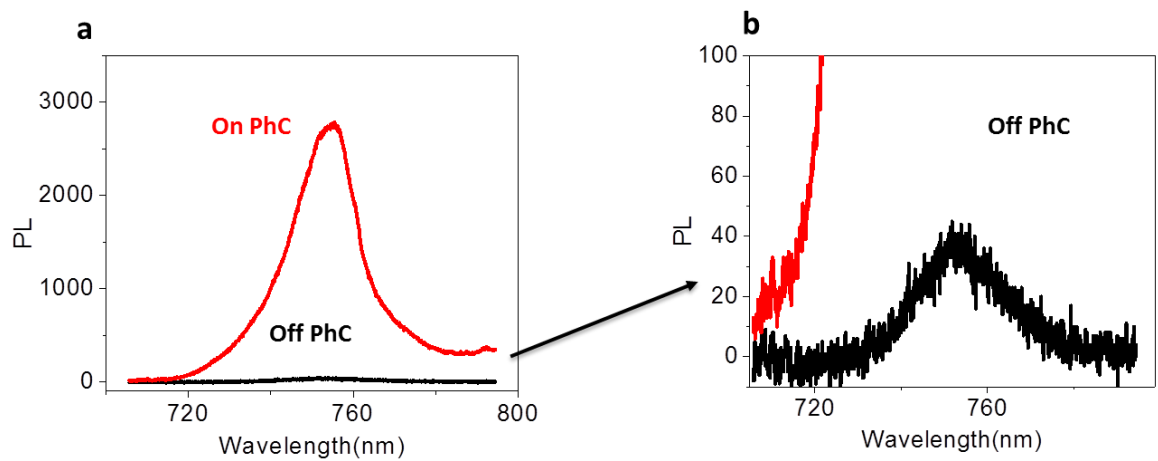

**Figure S1 | a,** Comparison of the PL spectra from on-PhC and off-PhC excitation in one of our devices, showing ~60X PL intensity enhancement for on-PhC. **b,** Zoom-in plot of the PL emission for off-PhC excitation.

## S2. Momentum resolved photoluminescence of monolayer WSe$_2$ on SiO$_2$ substrate

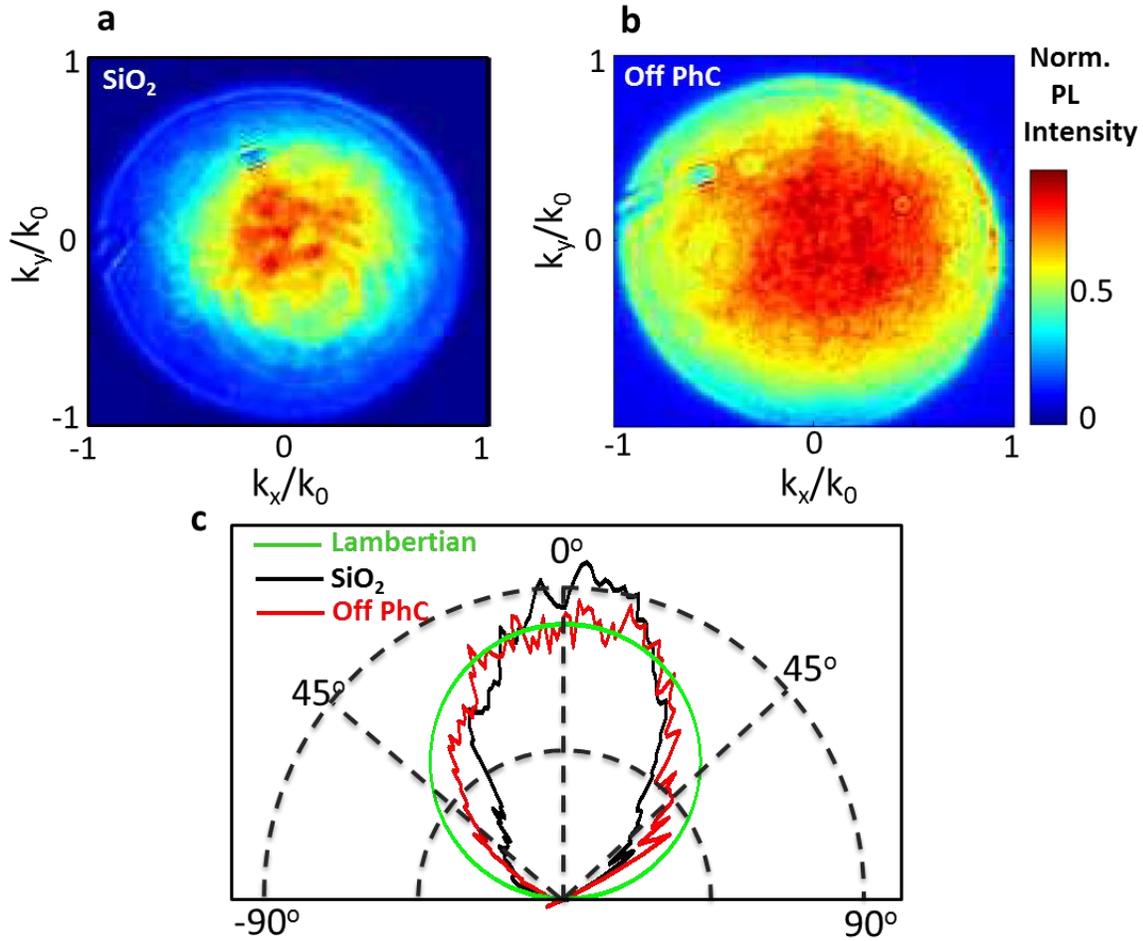

**Figure S2 | a,** Normalized PL intensity distribution over in-plane momentum space measured by momentum resolved microscopy, when the monolayer is placed on SiO$_2$ substrate. **b,** The same map for a sample on GaP substrate (re-plot Fig 3a for comparison). **c,** Polar plot of the two emission patterns at $k_x = 0$, compared with Lambertian emission. We can see that SiO$_2$ directs more light in the vertical direction.

## S3. Simulation of the emission distribution based on the diffraction grating effect

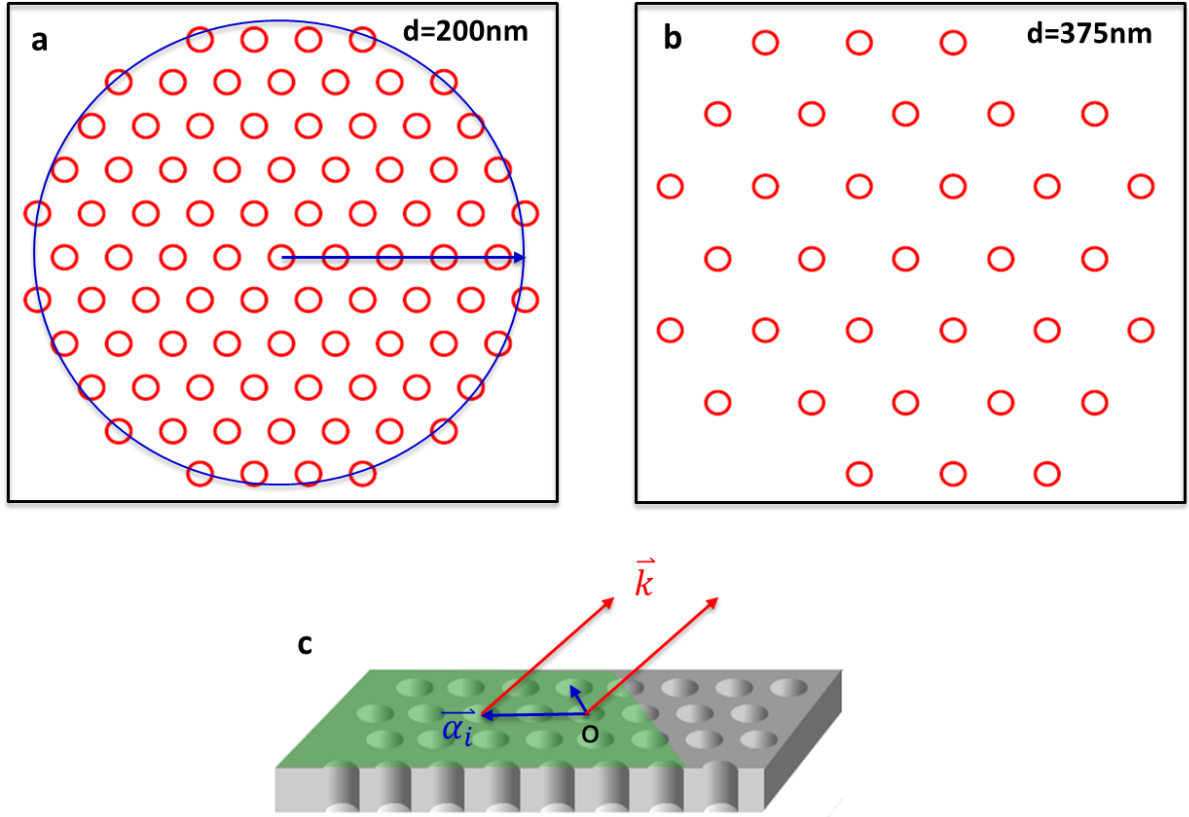

**Figure S3 | a,** Diffraction grating pattern used in theory for device A with 200nm lattice constant. The resulting far field emission pattern is shown in Fig. 3e in the main text. **b,** Diffraction pattern used for device B with 375nm lattice constant. The radius of the pattern is about one wavelength of the emitted light, as found to best match the observation. We found that such patterns reproduce the experimental observations for both devices. The diffraction grating effect is calculated by $I(\vec{k}) = |\sum_{\vec{\alpha_l}} A_l e^{i2\pi \vec{k}\cdot\vec{\alpha_l}/|\vec{k}|\lambda}|$, where $I(\vec{k})$ is the emission intensity with out-of-plane wave vector $\vec{k}$, $\vec{\alpha_l}$ is the in-plane lattice vector describing the corresponding $l^{th}$ hole and $A_l$ is the radiation intensity of the hole. Here we assume $A_l = 1$ for simplicity. The summation includes all the holes in patterns **a** and **b** for devices A and B, respectively. **c,** Cartoon illustrating the defined vectors and diffraction grating effect.

## S4. Cavity mode profile simulation

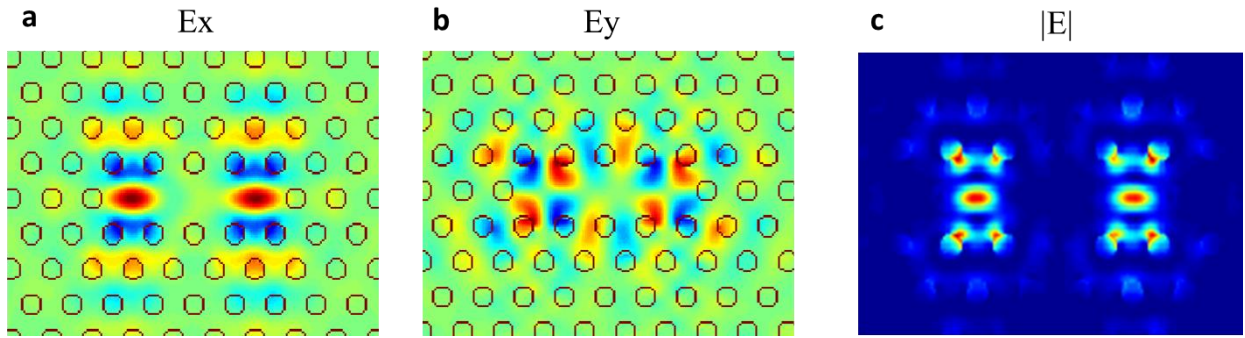

**Figure S4 |** There are multiple dipole-like modes in the L4 cavity with primary polarization in the *x*-direction. We present our simulated results of the electric field mode profiles (**a,** $E_x$; **b,** $E_y$; **c,** total electric field intensity), corresponding to the cavity emission peak, described in Fig. 2. In our simulation, the lattice constant $a = 200 nm$, hole radius $r = 0.25a$, membrane thickness is $180 nm$, emission wavelength is $744.4 nm$ and cavity quality factor $Q = 300$.